\documentstyle[sprocl,psfig]{article}

\def\Journal#1#2#3#4{{#1} {\bf #2}, #3 (#4)}

\def\be{\begin{equation}}
\def\ee{\end{equation}}
\def\bea{\begin{eqnarray}}
\def\eea{\end{eqnarray}}

\newcommand{\kms}	{km~s$^{-1}$}
\newcommand{\h}   	{$h^{-1}\,$~kpc}

\newcommand{\etal} 	{{et~al.}}

\newcommand{\lya}	{Ly$\alpha$}

\begin{document}


\title{The Detection of Ly$\alpha$ Absorption from Nine Nearby Galaxies}

\author{David V. Bowen}

\address{ Princeton University Observatory, Princeton, NJ 08544\\
E-mail: dvb@astro.princeton.edu} 

\author{Max Pettini}

\address{Institute of Astronomy, Madingley Rd.,
Cambridge CB3 0EZ, UK}

\author{J. Chris Blades}

\address{Space Telescope Science Institute, 3700 San Martin Drive, 
Baltimore, MD~21218}


\maketitle\abstracts{ We have used STIS aboard HST to search for \lya\
absorption in the outer regions of nine nearby ($cz <$ 6000~\kms )
galaxies using background QSOs and AGN as probes. 
The foreground galaxies are intercepted between 26 and 199~\h\ from
their centers, and
in all cases we detect \lya\ within $\pm500$~\kms\ of the galaxies'
systemic velocities. The intervening galaxies have a wide range of
luminosities, from $M_B = -17.1$ to $-20.0$, and reside in various
environments: half the galaxies are relatively isolated, the remainder
form parts of groups or clusters of varying richness.  The equivalent
widths of the \lya\ lines range from $0.08 - 0.68$~\AA\ and, with the
notable exception of absorption from one pair, correlate with
sightline separation in a way consistent with previously published
data, though the column densities derived from the lines do not. The
lack of correlation between line strength and galaxy luminosity or, in
particular, the environment of the galaxy suggests that the absorption
is not related to any individual galaxy, but arises in gas which
follows the same dark-matter structures that the galaxies inhabit.}

\section{Introduction}

The detection of $z<< 1$ \lya -forest absorption lines in the spectra
of QSOs observed by HST shortly after its launch
\cite{bah91,morr91,kp1} not only demonstrated the existence and
evolution of these tenuous neutral hydrogen (H~I) clouds over a
significant fraction of the age of the universe, but quickly sparked
an interest as to whether it might be possible to establish the
origin of the clouds themselves. Although generally thought to be
intergalactic at high redshift (because of their high rate of
incidence along a sightline, and their weak clustering~\cite{sarg80})
the remarkable success at detecting galaxies responsible for the
higher H~I column density Mg~II systems at low
redshift~\cite{jb1,BergBois,Stei94} supported the case for
investigating whether \lya\ absorption lines might also arise
in the halos of individual galaxies.

Mapping the galaxies around the sightline towards
3C~273~\cite{salz92,Morr93} produced little evidence for a direct
association between individual galaxies and \lya
-absorbers. Morris~\etal~\cite{Morr93} 
concluded that \lya\ clouds were not distributed
at random with respect to galaxies, nor did they cluster as strongly
as galaxies cluster with each other, and could only be associated with
galaxies on scales of $\sim 0.5 -1$~Mpc. However, in a study of six
different fields, Lanzetta~\etal~\cite{lbtw} found that {\it
a}) the majority of normal, luminous galaxies possess extended \lya
-absorbing halos or disks of radii $\sim 160$~\h , and {\it b})
between one and two thirds of all \lya\ lines arise in such
galaxies. Combined with other similar studies \cite{Stoc95,LeBr96} it
appeared that these disparate results might be reconciled if 
strong \lya\ lines were associated with galaxies, while
weak ones arose mainly in the intergalactic medium.

These initial results were soon re-evaluated in light of the rapid
development in hydrodynamical and semi-analytic simulations of how gas
behaves in hierarchical cold dark-matter structure formation.  These
simulations showed that gas should follow the same density fluctuations that
are gravitationally induced by the dark matter distributions,
resulting in a `web' of intersecting filaments and sheets of
gas. Analysis of artificial spectra, generated by shooting random
sightlines through the simulations, were extremely successful in
reproducing the observed properties of the high-redshift \lya
-forest~\cite{cen94,zhang95,hern96,mira96,bryan99}. In
particular, they showed that low column density lines are produced
predominantly in the filaments, while the higher column density lines
arise from denser gas in a virialized halo, i.e., the same regions in
which a high
galaxy density might be expected.

As the theoretical work continues, Chen~\etal\ (hereafter
CLWB)~\cite{Chen98} have now extended the original work of
Lanzetta~\etal\ and continue to find evidence for direct
galaxy-absorber association. They also find that the strength of the
absorption depends not only on impact parameter but also on galaxy
luminosity, suggesting a stronger link between galaxy and absorber.
Ortiz-Gil~\etal~\cite{Orti99} have associated {\it individual}
components within a complex \lya\ system with {\it individual}
galaxies from a group towards Q1545+2101, instead of an intragroup
medium (although the \lya\ lines are at the same redshift as the
QSO, so are not drawn from the same population as the lines
normally analyzed).  The simulators have also advanced their models to
$z\sim 0$, and have again been able to
reproduce many features of the observed \lya -forest~\cite{theuns,fortodd}.
Dav\'{e}~\etal~\cite{dave99} have used an algorithm designed to
identify clumps of gas and stars in their simulations which are likely
to correspond to galaxies, and impressively, have been able to
reproduce the correlation of line strength and impact parameter.

In a previous paper~\cite{bowen96} we used Archival HST FOS data to search
for \lya\ lines from present-day galaxies in order to better
understand whether \lya\ absorption arises in the halos of individual
galaxies. We found that, for lines stronger than $0.3$~\AA , {\it a})
nearby galaxies do not possess \lya -absorbing halos beyond 300~\h\ in
radius, and {\it b}) the covering factor of galaxies between 50 and
300~\h\ is $\sim 40$~\%. However, we found no correlation of \lya\
equivalent width with impact parameter or with galaxy luminosity, and
questioned whether the galaxies were indeed responsible for the
absorption lines. We instead concluded that our results supported the
picture emerging at the time that \lya\ lines arise in the sheets and
filaments discussed above.

Our analysis suffered from two major deficiencies, namely that we probed few
galaxies within the canonical 160~\h , and that we were restricted to looking
only for strong lines in the low resolution FOS data.  We sought to remedy these
deficiencies by obtaining more data with the STIS aboard HST, aiming to search
for {\it weak} lines {\it within} 160~\h\ of a nearby galaxy using the G140M
grating.  In this contribution, we outline some of the results obtained from
that program.  The experiment is not designed to address the origin of {\it all}
\lya\ absorbers, since we start by identifying a suitable galaxy and then search
for absorption from that galaxy. We do not seek to establish what fraction of
\lya\ absorbers arise in galaxy halos.

\begin{table}[t]
\caption{Probes observed by HST to search for \lya\ absorption from
foreground galaxies.}
\begin{tabular}{llcccc}
\hline\hline
		& Intervening	  & $v_{\rm{gal}}$  &   sep     & $M_B -$    & $W$    \\
Probe           & Galaxy          & {\footnotesize (\kms )}         & {\footnotesize ($h^{-1}$ kpc)}   & $5\log h$  & (\AA ) \\
\hline
Mrk 1048        & NGC 988         & 1504            & 158     &	$-20.0$ & $0.11\pm0.01$ \\
PKS 1004+130    & UGC 5454        & 2792            & 84      & $-17.9$ & $0.68\pm0.05$ \\
ESO 438$-$G009  & UGCA 226        & 1507            & 110     & $-17.1$ & $0.36\pm0.06$ \\
MCG+10$-16-$111 & NGC 3613        & 1987            & 26      & $-19.8$ & $0.54\pm0.02$ \\
		& NGC 3619        & 1542            & 85      & $-18.5$ & $0.59\pm0.02$ \\
PG 1149$-$110   & NGC 3942        & 3696            & 92      & $-19.1$ & $0.40\pm0.06$ \\
Q1341+258       & G1341+2555      & 5802            & 31      & $-18.0$ & $0.08\pm0.02$ \\
Q1831+731       & NGC 6654        & 1821            & 143     & $-18.9$ & $0.11\pm0.02$ \\
		& NGC 6654A       & 1558            & 199     & $-18.5$ & $0.10\pm0.01$ \\
\hline
\end{tabular}
\end{table}

\section{HST Observations}

In order to produce a sample of QSO-galaxy pairs which could be
observed with HST, we cross-correlated the Third Reference Catalogue
of Bright Galaxies~\cite{RC3} with version 7 of the QSO/AGN catalog of
V\'{e}ron-Cetty \& V\'{e}ron (1996). We chose galaxies with velocities
$ > 1300$~\kms , since absorption below these values would likely be
lost in the damped \lya\ absorption profile from the Milky Way. The
final group of QSO-galaxy pairs successfully observed by HST is listed
in Table~1. Half of the galaxies are relatively isolated, while the
other half are found in groups of various richness.  In Figures 1 \& 2
we show two examples of the fields studied, the poor group towards
PKS~1004+130 and the rich group towards MCG+10$-$16$-111$. In Figure~3
we show three representative spectra: two are of the QSOs shown in the
first two figures, while the third is of Q1341+258 which probes the
halo of G1341+2555 at the small separation of 31~\h .

We detect \lya\ lines within a few hundred \kms\ from all nine
galaxies. This suggests that galaxies in the local universe are indeed
surrounded by low column density H~I with $\log
N$(H~I)$\sim 13-15$ at radii of $26-200$~\h . The ubiquity of the
detections suggests a high covering factor --- 
$\sim 100$~\% at these column
densities and radii. A plot of equivalent width versus impact
parameter shows a weak correlation (not shown herein), consistent with
CLWB's results, although column density vs.~impact parameter is
uncorrelated.  There also appears to be no dependency of line strength
with any other parameters such as galaxy magnitude or morphology.

The question remains, therefore, whether the neutral hydrogen detected has
anything to do with the galaxy itself. In light of the success that the
hydrodynamical simulations have had embedding galaxies in sheets of H~I (\S1),
does the detection of gas near a galaxy reflect anything more than the fact that
gas and galaxy share the same gravitational potential? Our data show little
evidence for individual galaxies producing or influencing the halos around
themselves: for example, the detection of at least five individual \lya\
components towards PKS 1004+130 at the velocity of UGC~5454 and a companion Low
Surface Brightness galaxy LSBC~D637$-$18 (Fig.~1), spanning 740~\kms , is hard to
understand as arising in two overlapping halos 84 and 138~\h\ from the QSO
sightline. On the other hand, intragroup gas could be expected to show a broad
velocity spread due to the velocity dispersion of the group. Further, the detection of
weak absorption from the isolated galaxy G1341+2555 towards Q1341+258 (Fig.~3,
bottom panel) when the impact parameter is only 31~\h\ is very rare---there
are no such cases of such small values of equivalent width for such a small
impact parameter in CLWB's sample. Finally, the strong \lya\ lines detected towards
MCG+10$-$16$-111$ (Fig.~2) are coincident in velocity with a strong over-density
of galaxies within 2~$h^{-1}$~Mpc of the sightline, again suggesting that
intragroup gas is probably responsible for the absorption. It seems likely
therefore that the H~I we detect---and indeed, the galaxies we chose to
probe---both reflect the underlying gravitational fluctuations, as the
simulations predict.

\bigskip


{\small Support for
this work was provided through grant GO-08316.01 from the
Space Telescope Science Institute, which is operated by 
the Association of Universities for Research in Astronomy, Inc.,
under NASA contract NAS5-26555.}


\begin{figure}
\vspace*{-4cm}\centerline{\psfig
{figure=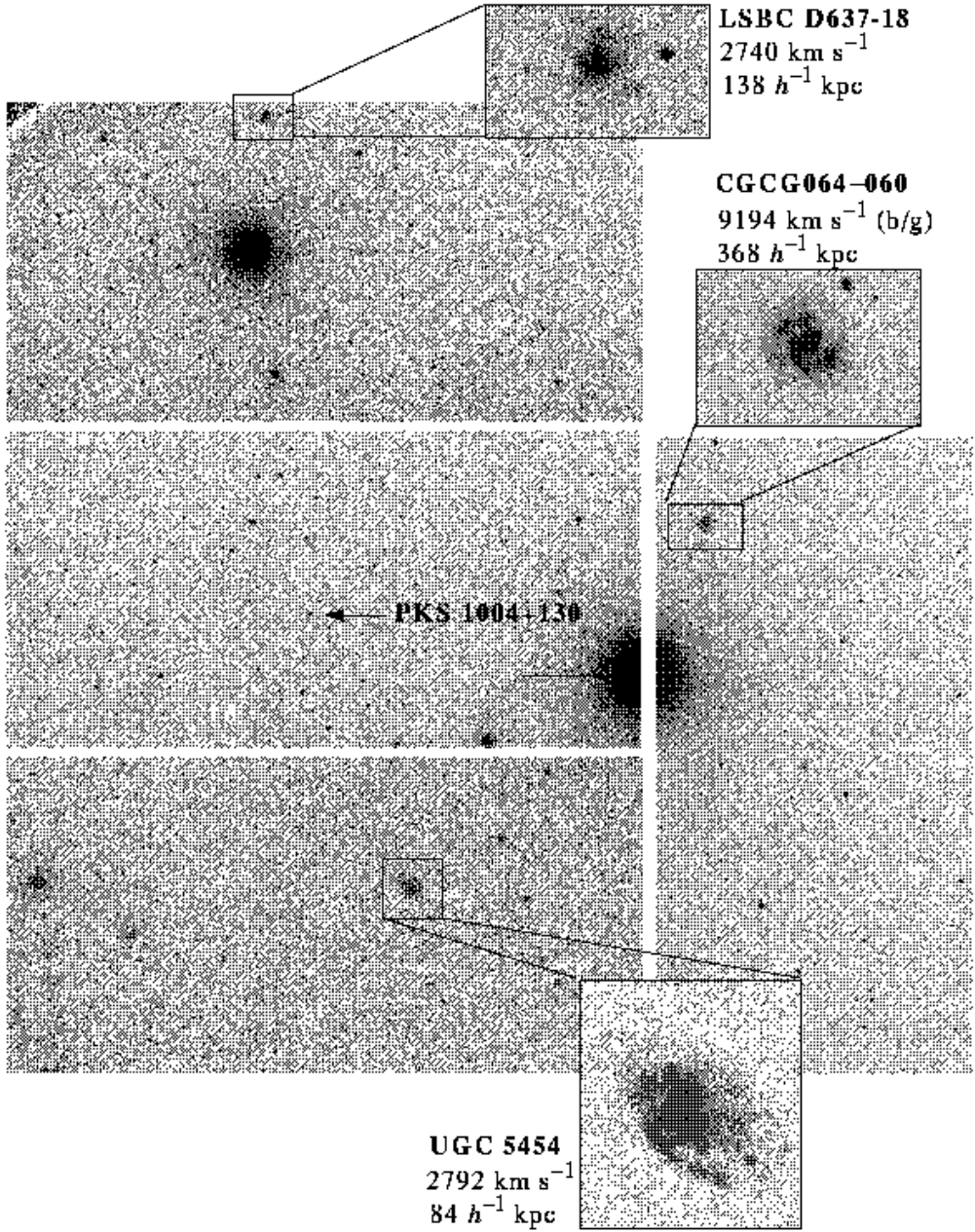,height=18cm,angle=0}}
\caption{Reproduction of an Isaac Newton Telescope {\it Wide Field
Camera} 
image of the field
around the QSO PKS~1004+130 ($z=0.240$). The field contains the dwarf
galaxy UGC~5454 and an LSB galaxy LSBC~D637$-$18. Below each designation, the
galaxy's velocity and separation from the QSO sightline is
given. For scale, the separation between the QSO and UGC~5454 is 10.5
arcmins. \lya\ absorption is found to arise at the velocity of
UGC~5454 \& LSBC~D637$-$18 (Fig.~3). The absorption is complex, consisting of at least
five individual components, spanning a velocity interval of 740~\kms
. Such strong \& complex absorption is unlikely to arise from the halo
of UGC~5454 \& LSBC~D637$-$18 alone, but probably from intragroup gas.}
\end{figure}

\psfull

\begin{figure}
\vspace*{-5cm}\centerline{\psfig
{figure=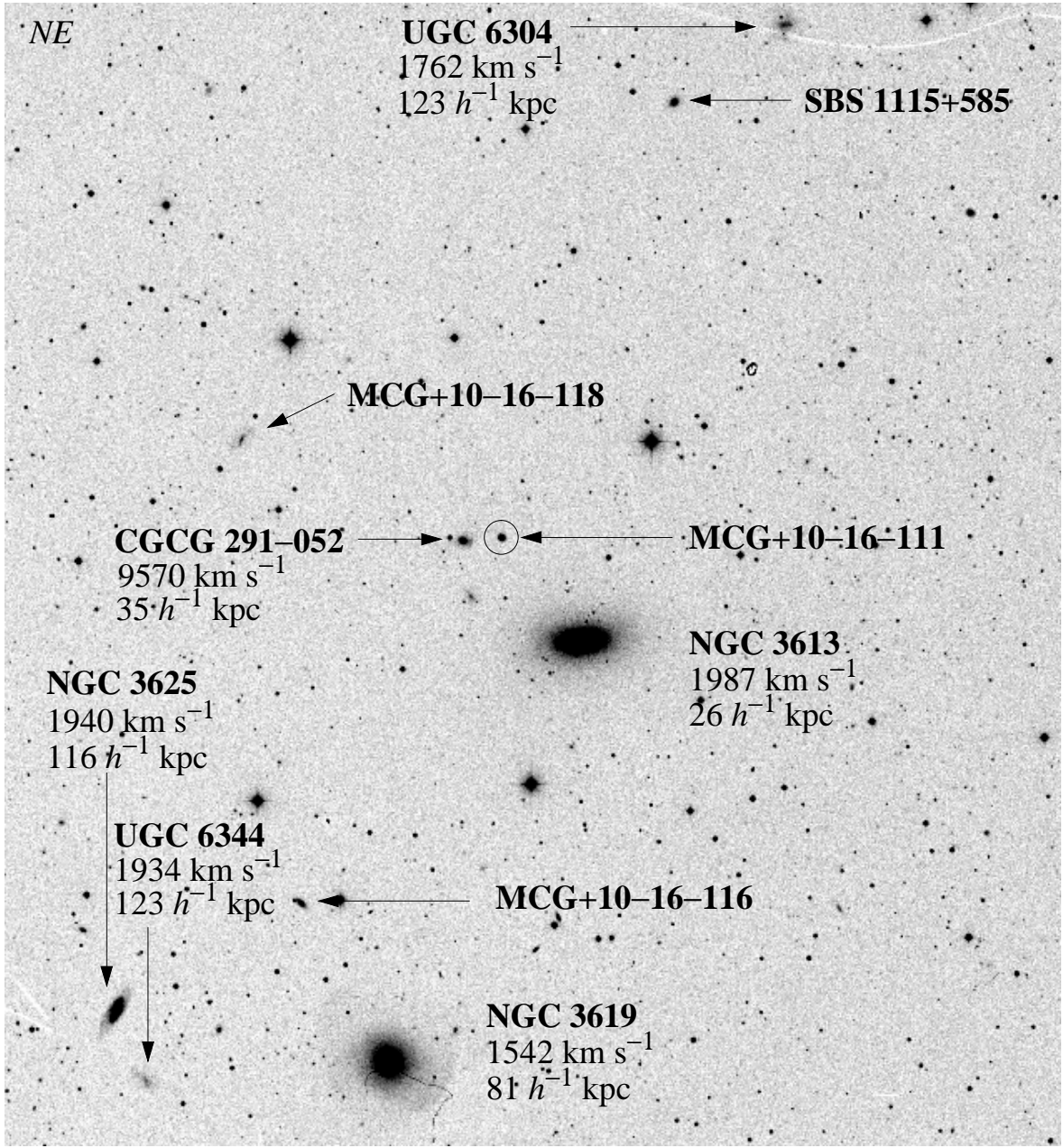,height=22cm,angle=0}}
\vspace*{-3cm}\caption{Reproduction of the STScI Digitized Sky Survey centered around
MCG+10$-$16$-111$ ($z=0.027$). The field is dominated by two
bright galaxies, NGC~3613 \& NGC~3619, although many fainter galaxies
lie within radii of less than 150~\h. Below each designation, the
galaxy's velocity and separation from the probe's sightline is again
given, although redshifts are not available for all galaxies identified. For
scale, 
the separation between the MCG+10$-$16$-111$ (circled) and NGC~3619 is 18.2 arcmins.
Strong \lya\ absorption is detected at velocities of many of
these galaxies (Fig.~3), although whether a single galaxy gives rise to the
absorption is unclear.}
\end{figure}

\begin{figure}
\vspace*{-4.5cm}\centerline{\psfig
{figure=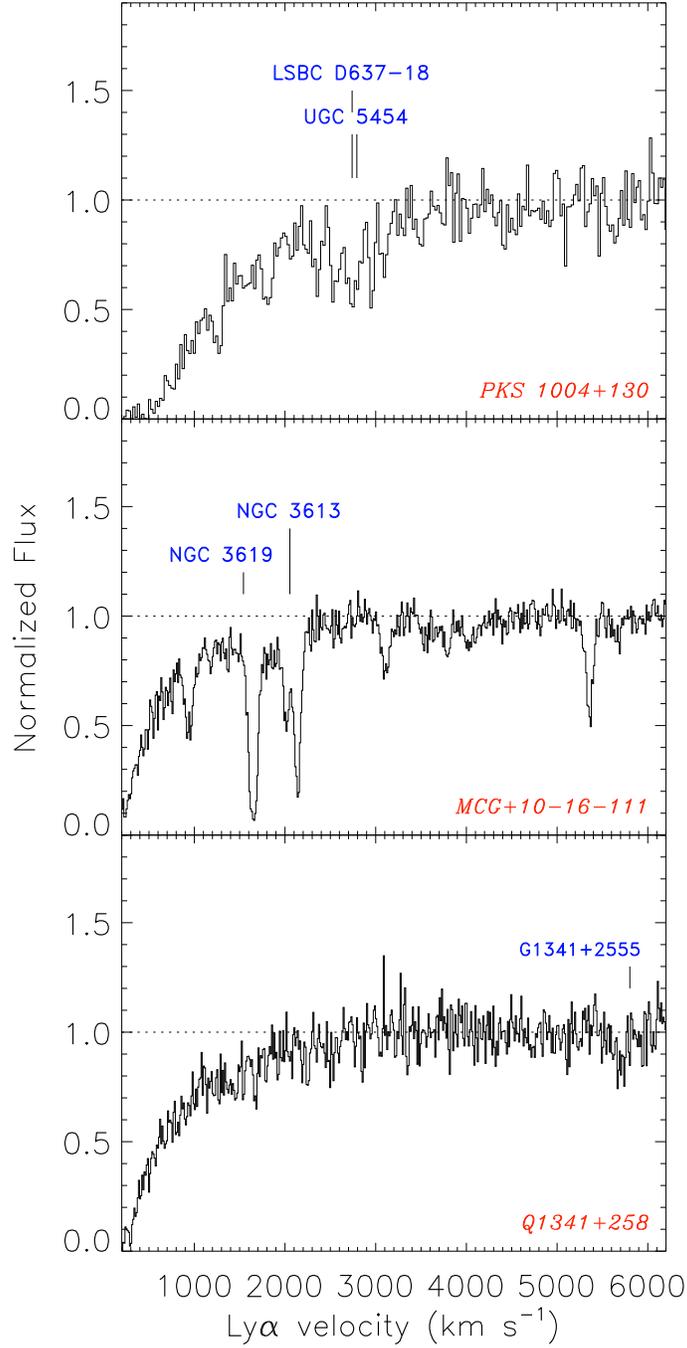,height=19cm,angle=0}}
\vspace*{-0.5cm}\caption{Three HST spectra from our survey. 
The strong decrease in flux at $v < 1000$~\kms\ 
is due to the damped \lya\ profile arising from absorption by Milky Way
H~I. Velocities of the galaxies close to the probe
sightline are labeled.
The first two spectra
show the absorption from galaxies in the fields presented in
Figs. 1 \& 2. The spectrum of PKS~1004+130 is binned 2x that of the other
spectra due to its low S/N. The third spectrum shows extremely weak absorption  towards
Q1341+258 
from a foreground galaxy only 31~\h\ from the sightline. 
(The line may be blended with Galactic N~V~$\lambda 1238$, which would make the
intrinsic strength of the \lya\ line even weaker.)
Such weak  absorption so close to a galaxy is extremely unusual compared to
the equivalent widths found for galaxies at similar separations by CLWB. }
\end{figure}

\end{document}